\documentclass[10pt]{article}

\usepackage{natbib}

\usepackage{graphicx}

\usepackage{natbib}
\usepackage{amsmath}
\usepackage{amssymb}
\usepackage{psfrag}
\usepackage{stmaryrd}
\usepackage{bm}

\begin{document}
%-------------------------------------------------------------------
% Defines (OG)
%-------------------------------------------------------------------
%
\newcommand{\gradn}{{\nabla}}
\newcommand{\disc}[1]{\left\llbracket {#1} \right\rrbracket}
\newcommand{\ave}[1]{\left< {#1} \right>}

\newcommand{\brac}[1]{\left( {#1} \right)}
\newcommand{\bracc}[1]{\left\{ {#1} \right\}}
\newcommand{\bracs}[1]{\left( {#1} \right)^{\text{s}}}
\newcommand{\bs}[1]{\bm{#1}}
\newcommand{\vect}{\bs}
\newcommand{\pd}{\partial}
\newcommand{\dif}{ \ d}
\newcommand{\dotp}{\bs{\cdot}}
\newcommand{\ip}{\mathbf{:}}
\newcommand{\grad}{{\nabla}^{\rm s}}
\newcommand{\eps}{\varepsilon}
\newcommand{\strain}{\bs{\eps}}
\newcommand{\stress}{\bs{\sigma}}

\renewcommand{\Bar}{\overline}
\renewcommand{\Tilde}{\widetilde}

%begin paper here

%\runningheads{L. Molari, G. N. Wells, K. Garikipati, F. Ubertini}{A
%Continuous/Discontinuous Galerkin Method for Gradient Damage}
%\begin{frontmatter}
\title{A discontinuous Galerkin method for strain gradient-dependent damage:
    Study of interpolations, convergence and two dimensional problems}

\author{Luisa Molari
\\
DISTART, Universit\`{a} di Bologna, \\
    Viale Risorgimento 2, 40136 Bologna, Italy
\\[20pt]
Garth N. Wells
\\
Faculty of Civil Engineering and
    Geosciences, Delft University of Technology, \\
    Stevinweg~1, 2628 CN Delft, The Netherlands
\\[20pt]
Krishna Garikipati
\\
Department of Mechanical Engineering, University of Michigan, \\
            Ann Arbor, Michigan 48109,  USA
\\[20pt]
Francesco Ubertini
\\
DISTART, Universit\`{a} di Bologna, \\
    Viale Risorgimento 2, 40136 Bologna, Italy}

\maketitle

\begin{abstract} \footnotesize
A discontinuous Galerkin method has been developed for strain gradient-dependent
damage. The strength of this method lies in the fact that it allows the use of
$C^0$ interpolation functions for continuum theories involving higher-order
derivatives, while in a conventional framework at least $C^1$ interpolations are
required. The discontinuous Galerkin formulation thereby offers significant
potential for engineering computations with strain gradient-dependent models.
When using basis functions with a low degree of continuity, jump conditions
arise at element edges which are incorporated in the weak form. In addition to
the formulation itself, a detailed study of the convergence properties of the
method for various element types is presented, an error analysis is undertaken,
and the method is also shown to work in two dimensions.
\end{abstract}

%\begin{keyword} \footnotesize
%    Strain gradient theory,
%    mixed methods, stabilised methods, damage.
%\end{keyword}
%\end{frontmatter}

%\tableofcontents

%\clearpage

\section{Introduction}

The nucleation and growth of cracks can be described by continuum damage
mechanics. In this approach, an internal variable is included in the
constitutive law to represent the evolution of microstructural
damage~\citep{Kachanov}. The material moduli degrade with the increase of this
parameter, which could be either scalar or tensorial in character. A scalar
internal variable is commonly used to model damage degradation for cases in
which the material remains isotropic. A tensorial form of the internal variable
is used for the anisotropic case.

Damage degradation can manifest itself in progressive material softening, for
which reason numerical results based upon classical continuum mechanics are
characterised by a pathological mesh dependence \citep{Willam:84,Bazant:86}. As
the material softens, deformations localise within bands of finite thickness,
which is viewed as the smearing out of a crack \citep{Rashid:68,bazant:1983}.
However, the width of bands is intimately related to the element size. As the
mesh is refined the band width decreases, tending to zero in the limit. The
result of this is seen in load-displacement response with slopes that are
increasingly negative with mesh refinement in the softening regime.
The origin of this pathology lies in the loss of ellipticity of the
material tangent modulus tensor for softening inelastic continuum models in the
absence of rate effects \citep{Rice:76}.

The last two decades have witnessed a great deal of work in `regularised'
continuum models to avoid the loss of ellipticity in the presence of
material softening. Among these are the strain gradient models
\citep{aifantis:1984, Coleman:85, Triantafyllidis:86, Peerlings:96}, nonlocal
models
\citep{Baz1, devree, borino:2003}, localisation limiters \citep{lar}, and
Cosserat
models \citep{DeBorstSluys:91}, to identify but a fraction of a vast body of
literature.

By various techniques, each of these models introduces an intrinsic length scale
to the continuum theory. When the material softens, the localisation band width
is controlled by this length scale, rather than the element size. As a result,
the mesh size-dependency, described above, is eliminated.

In this paper we aim to address numerical issues associated with some strain
gradient models. This class of models involves strains and strain gradients as
kinematic terms. In some models work conjugate couple stresses are also
identified. Dimensional considerations lead to the introduction of intrinsic
length scales associated with the strain gradients. The presence of strain
gradient terms alters the mathematical character of the governing differential
equations, elevating them to a higher order (at least locally). A numerical
complication arises from the higher order character of the governing
differential equations, as standard $C^{0}$ interpolation functions often do
not furnish sufficient continuity.

The obvious approach, that is to employ interpolation functions with
higher-order continuity ($C^{1}$ and higher), has not proven fruitful. Finite
element methods with these interpolations are expensive and difficult to
construct.
There exists some work with mixed formulations for this purpose
\citep{Shuetal:99, zervos:2001}. However, the very simplest two-dimensional
elements involve up to 28 degrees of freedom, placing them beyond the realm of
practicability.
Some authors  have applied element-free Galerkin methods, on the basis of the
arbitrary degree of continuity that they can provide~\citep{Askes:2000}.
However, this class of methods introduces complications in the  application of
boundary conditions and is somewhat less efficient than the finite element
method, and  its implementation involves a reformulation of the dominant
finite element architecture, making its widespread use less attractive.
In addition, just as high-order continuity can be difficult to achieve,
excessive continuity can also be a drawback for many models, particularly at
internal boundaries where the order of the governing differential equation
changes.

In this work we seek to expand upon recent developments in discontinuous
Galerkin methods for strain gradient-dependent theories \citep{Engeletal:2002,
Wellsetal:2003}. The great advantage of this class of methods is the ability to
use $C^{0}$ interpolation functions for the displacement when solving continuum
theories involving higher-order derivatives. In cases where other considerations
make it advantageous to also interpolate strain-like quantities, the approach
can allow interpolated fields that are discontinuous across elements
\citep{Wellsetal:2003}.

The content of this paper is organised as follows. The strain gradient damage
model of interest is described in Section~\ref{sect2}, with special attention
paid to the boundary conditions. The Galerkin formulation for this model is
presented in Section~\ref{sect3}. The convergence behaviour of the model in one
dimension is examined both analytically and numerically in Section~\ref{sect4},
and the section is concluded with one- and two-dimensional examples.
Conclusions are then drawn in Section~\ref{sect5}.

\section{Gradient damage}
\label{sect2}

Consider a body $\Omega$, which is an open subset in $\mathbb{R}^{n}$, where
$n$ is the spatial dimension. The boundary of the body is denoted by $\Gamma =
\pd \Omega$, and the outward unit normal to $\Gamma$ is denoted by~$\vect{n}$.
We consider quasi-static equilibrium of the body, which is governed by
\begin{align}
    \gradn \cdot \bs{\sigma} + \vect{f} &= \vect{0} &&{\rm in}\; \Omega,
                                                    \label{goveq}\\
    \bs{\sigma} \vect{n}                &= \vect{h} &&{\rm on}\; \Gamma_{h},
                                                    \label{stressbc}\\
    \vect{u}                            &= \vect{g} &&{\rm on}\; \Gamma_{g},
                                                    \label{dispbc}
    \end{align}
where $\stress$ is the stress tensor, $\vect{f}$ is the body force, $\vect{h}$
is the prescribed traction on the boundary $\Gamma_{h}$ and $\vect{g}$ is
the prescribed displacement on the boundary $\Gamma_{g}$. The boundary $\Gamma$
is partitioned such that $\Gamma = \overline{\Gamma_{g} \cup
\Gamma_{h}}$ and $\Gamma_{g} \cap \Gamma_{h} = \emptyset$. For isotropic
damage, it is sufficient to introduce a single scalar damage variable, $D$,
satisfying $0\leq D\leq 1$. Considering that $D = 0$ implies no damage and $D =
1$ corresponds to a completely damaged material point, the stress-strain
relation is of the form
    \begin{equation}
        \bs{\sigma}  =\brac{1-D} \mathbb{C} \ip \strain,
        \label{stress-strain}
    \end{equation}
where $\mathbb{C}$ is the elasticity tensor and $\strain$
is the strain tensor, $\strain = \grad \vect{u}$. The scalar damage variable,
$D$, is related to a scalar strain measure, $\bar{\eps}$, through the
Kuhn-Tucker conditions:
    \begin{equation}
        \Bar{\eps}-\kappa\le 0, \ \dot{\kappa}\ge 0, \
                \dot{\kappa}(\bar{\eps}-\kappa)=0,
        \label{kuhn}
    \end{equation}
where $D = D(\kappa)$.
The scalar strain measure $\bar{\eps}$ of interest here arises from a so-called
explicit gradient damage formulation. In this approach, the strain measure is
defined to be
    \begin{equation}
        \Bar{\eps} = \eps_{\rm eq} + c^2\nabla^{2}\eps_{\rm eq},
        \label{epsbar}
    \end{equation}
where $c$ is an intrinsic length scale, and $\eps_{\rm eq}$ is a suitably chosen
invariant of the local strain tensor.

For material points at which damage is developing, the governing equation is
nonlinear and fourth-order in the displacement. In undamaged or unloading
regions ($\Dot{\kappa} = 0$), the governing equation is linear and second-order.
Recognising the fourth-order form of the differential equation (at least in
sub-regions of $\Omega$) leads to the question of appropriate boundary
conditions. Possible boundary conditions (in addition to the usual
boundary conditions on the displacement and traction) include:
    \begin{align}
        \eps_{\rm eq}                           &= q    && {\rm on} \ \Gamma_{\eps}, \\
        \gradn \eps_{\rm eq}\cdot \vect{n}_{d}  &= r    && {\rm on} \
                                                            \Gamma_{\eps^{\prime}},  \\
        \brac{\gradn \brac{ \gradn  \eps_{\rm eq} } \vect{n}_{d} }
                            \cdot \vect{n}_{d} &= s     &&
                {\rm on} \ \Gamma_{\eps^{\prime\prime}},
        \label{hobc}
    \end{align}
where
$\Bar{\Gamma_{\eps} \cup \Gamma_{\eps^\prime}  \cup \Gamma_{\eps^{\prime
\prime}}} = \Gamma_{d}$,
$\Gamma_{\eps} \cap \Gamma_{\eps^{\prime}} = \emptyset$,
$\Gamma_{\eps} \cap \Gamma_{\eps^{\prime\prime}} = \emptyset$,
$\Gamma_{\eps^{\prime}} \cap \Gamma_{\eps^{\prime\prime}} = \emptyset$,
and $\vect{n}_{d}$ indicates the unit outward normal to the boundary of the
damaged domain, $\Gamma_{d}$. These boundary conditions supplement the standard
displacement and traction boundary conditions on~$\Gamma$.
If the
entire body $\Omega$ is damaging, then $\Gamma_{d} = \Gamma$. Commonly however,
damage is localised. In this case interface conditions arise on the boundaries
which are internal to $\Omega$ (the boundaries to sub-domains where
$\dot{\kappa} > 0$ and $\dot{\kappa} = 0$ meet). When the body force is smooth,
continuity of $\vect{u}$ and $\stress \vect{n}_{d}$ are natural continuity
conditions. At the interface between damaging and elastic regions, an additional
condition is required on the damage domain. The need for extra boundary
conditions is made particularly clear by the analytical solution presented in
later in Section~\ref{sec:exact}.
In the adopted formulation, this
extra condition will come from implied continuity of  $\gradn \eps_{\rm eq}
\cdot \vect{n}_{d}$ across the damage-elastic boundary.

%-------------------------------------------------------

\section{Galerkin formulation}
\label{sect3}
    In developing a weak formulation for eventual finite element solution, the
equilibrium equation~\eqref{goveq} and the equation for $\Bar{\eps}$
\eqref{epsbar}  are considered separately. The nonlinear fourth-order equation
resulting from insertion of the constitutive equations into the equilibrium
equation could potentially be cast in a weak form, with integration by parts
applied twice. The formulation would be specific to the chosen dependence of
damage on $\kappa$, which is typically complex.
Furthermore, the complexities associated with a moving elastic-damage boundary
(which is the interface between second- and fourth-order sub-domains) would be
significant. Hence, it is convenient to treat the two equations separately.

    The body $\Omega$ is partitioned into $n_{\rm el}$ non-overlapping elements
$\Omega_{e}$ such that
    \begin{equation}
        \Bar{\Omega} = \bigcup_{e=1}^{n_{\rm el}} \Bar{\Omega}_{e},
    \end{equation}
where $\Bar{\Omega}_{e}$ is a closed set (i.e., it includes the boundary of the
element). The elements $\Omega_{e}$ (which are open sets) satisfy the standard
requirements for a finite  element partition. A domain $\Tilde{\Omega}$ is also
defined:
    \begin{equation}
        \Tilde{\Omega} = \bigcup_{e=1}^{n_{\rm  el}} \Omega_{e},
    \end{equation}
where $\Tilde{\Omega}$ does not include element boundaries. It is also useful to
define the `interior' boundary $\Tilde{\Gamma}$,
    \begin{equation}
        \Tilde{\Gamma} = \bigcup_{i=1}^{n_{b}} \Gamma_{i},
    \end{equation}
where $\Gamma_{i}$ is the $i$th interior element boundary and $n_{b}$ is the
number of internal inter-element boundaries.

    Consider now the function spaces $\mathcal{S}^{h}$, $\mathcal{V}^{h}$ and
$\mathcal{W}^{h}$,
    \begin{align}
        \mathcal{S}^{h} &=\bracc{u^{h}_{i} \in H^{1}\brac{\Omega} \ \left| \
                u_{i}^{h}|_{\Omega_{e}} \in P_{k_{1}}\brac{\Omega_{e}}
                        \forall \ e,
                        \ u_{i} = g_{i} \ {\rm on}
                        \ \Gamma_{g} \right. }  \label{eqn:disp_trial},          \\
        \mathcal{V}^{h} &=\bracc{w^{h}_{i} \in H^{1}\brac{\Omega} \ \left| \
                w_{i}^{h}|_{\Omega_{e}} \in P_{k_{1}}\brac{\Omega_{e}}
                                    \forall \ e, \ w_{i} = 0 \ {\rm on}
                        \ \Gamma_{g} \right. },  \label{eqn:disp_test}      \\
        \mathcal{W}^{h} &=\bracc{q^{h} \in L^{2}\brac{\Omega} \ \left| \
                q^{h}|_{\Omega_{e}} \in P_{k_{2}}\brac{\Omega_{e}} \forall \ e
                \right. },       \label{eqn:space_q}
    \end{align}
where $P_{k}$ represents the space of polynomial finite element shape functions
of order $k$. The spaces $\mathcal{S}^{h}$ and $\mathcal{V}^{h}$ represent
usual, $C^{0}$ continuous finite element shape functions. Note that the space
$\mathcal{W}^{h}$ contains discontinuous functions.

\subsection{Standard Galerkin weak form}
    The standard, continuous Galerkin problem for the equilibrium equation
\eqref{goveq} is of the form: given $\Bar{\eps}^{h} \in \mathcal{W}^{h}$, find
$\vect{u}^h \in \brac{\mathcal{S}^{h}}^{n}$ such that
    \begin{align}
        \int_{\Omega} \grad \vect{w}^{h} \ip
            \brac{1 - D\brac{\Bar{\eps}^{h}}} \mathbb{C} \ip
                    \grad \vect{u}^{h} \dif \Omega
         - \int_{\Gamma_{h}} \vect{w}^{h} \cdot \vect{h} \dif \Gamma = 0
            && \forall \ \vect{w}^{h} \in \brac{\mathcal{V}^{h}}^{n},
\label{eqn:equil_cont}
    \end{align}
where it was already assumed that $\vect{u}^{h}$ is $C^{0}$ continuous (see
equation \eqref{eqn:disp_trial}).
A second Galerkin problem is constructed to solve for $\Bar{\eps}$ (equation
(\ref{epsbar})). It consists of: given $\vect{u}^{h} \in
\brac{\mathcal{S}^{h}}^{n}$, find $\Bar{\eps}^{h} \in \mathcal{W}^{h}$ such that
    \begin{multline}
        \int_{\Omega} q^{h} \Bar{\eps}^{h} \dif \Omega
            -\int_{\Omega} q^{h} \eps_\mathrm{eq}^{h} \dif \Omega
            +\int_{\Omega} \gradn q^{h}\cdot c^{2} \gradn \eps_\mathrm{eq}^{h}
                \dif \Omega
                \\
            -\int_{\Gamma} q^{h}
            c^2\gradn\eps_\mathrm{eq}^{h} \cdot \vect{n}
                \dif \Gamma
            = 0 \quad \  \forall \ q^{h} \in \mathcal{W}^{h}.
    \label{eqn:bar_eps_weak}
    \end{multline}
Recall that discontinuities in $q^{h}$ and $\Bar{\eps}^{h}$ across
$\Tilde{\Gamma}$ are permitted. At this stage,
the only boundary conditions implied by the formulation are on the displacement
field (by construction) and the traction. Discussion regarding the
enforcement of `non-standard' boundary conditions is delayed until the following
section.

    Two problems exist in the preceding Galerkin formulation. The first is
that the weight function $q^{h}$ can be discontinuous, hence $\gradn
q^{h}$ is not necessarily square-integrable on $\Omega$. This problem can be
circumvented easily by requiring $C^{0}$ continuity of the functions
in~$\mathcal{W}^{h}$. The second problem, which is less easily solved, is that
$\eps_\mathrm{eq}^{h}$ is computed from $\grad \vect{u}^{h}$. Therefore,
calculating $\gradn \eps_\mathrm{eq}^{h}$ everywhere in $\Omega$ requires that
the displacement field $\vect{u}^{h}$ be $C^{1}$ continuous if singularities
are to be avoided on $\Tilde{\Gamma}$. However, since $\vect{u}^{h} \in
H^{1} \brac{\Omega}$ (see equation \eqref{eqn:disp_trial}), it is
not necessarily $C^{1}$ continuous.

\subsection{Discontinuous Galerkin form}
    The approach advocated here avoids the need for $C^{1}$ continuity of
the displacement field by imposing the required degree of continuity in a weak
sense. Before proceeding with the formulation, jump and averaging operations
are defined. The jump in a field $\vect{a}$ across a
surface (which is associated with a body) is given by:
    \begin{equation}
        \disc{\vect{a}} = \vect{a}_{1} \cdot \vect{n}_{1} +
        \vect{a}_{2} \cdot \vect{n}_{2},
    \end{equation}
where the subscripts denote the side of the surface and $\vect{n}$ is the
outward unit normal vector ($\vect{n}_{1} = - \vect{n}_{2}$ in the
geometrically linear case).
%This definition is convenient as it avoids the
%introduction of  `+' and `-' denotations of the sides of a surface. This is
%particularly useful for arbitrarily-oriented surfaces in two and three
%dimensions.
The average of a field $\vect{a}$ across a surface is given by:
    \begin{equation}
        \ave{\vect{a}} = \frac{\vect{a}_{1} + \vect{a}_{2}}{2}.
    \end{equation}

    Consider now the problem \citep{Wellsetal:2003}: given $\vect{u}^{h} \in
\brac{\mathcal{S}^{h}}^{n}$, find $\Bar{\eps}^{h} \in \mathcal{W}^{h}$ such that
    \begin{multline}
        \int_{\Omega} q^{h} \Bar{\eps}^{h} \dif \Omega
            -\int_{\Omega} q^{h} \eps_{\mathrm{eq}}^{h} \dif \Omega
            +\int_{\Tilde{\Omega}} \gradn q^{h} \cdot  c^{2} \gradn
            \eps_{\mathrm{eq}} ^{h} \dif \Omega
            -\int_{\Gamma} q^{h} c^2\gradn \eps^{n}_{\rm eq} \cdot \vect{n}
\dif \Gamma
            \\
            -\int_{\Tilde{\Gamma}} \disc{q^{h}} \cdot
                    c^{2} \ave{\gradn \eps_{\mathrm{eq}}^{h}} \dif \Gamma
            -\int_{\Tilde{\Gamma}} \ave{\gradn q^{h}} \cdot
                        c^{2}\disc{\eps_{\mathrm{eq}}^{h}} \dif \Gamma
                        \\
            + \int_{\Tilde{\Gamma}} \frac{\alpha c^{2}}{h} \disc{q^{h}} \cdot
                    {\disc{\eps_{\mathrm{eq}}^{h}}} \dif \Gamma
            =
            0 \quad \ \forall \ q^{h} \in \mathcal{W}^{h},
        \label{eqn:eps_weak}
    \end{multline}
where $\alpha$ is a penalty-like parameter, and $h$ is the element
dimension. No gradients of $\eps_{\mathrm{eq}}^{h}$ or $q^{h}$ appear in terms
integrated over $\Omega$ (which includes interior boundaries) in equation
(\ref{eqn:eps_weak}), hence the continuity requirements on the spaces
$\mathcal{S}^{h}$ and $\mathcal{W}^{h}$ are sufficient.

    Equation~\eqref{eqn:eps_weak} resembles the `interior penalty' method, which
belongs to the discontinuous Galerkin family of methods~\citep{Arnold:2002}.
Terms have been added to the weak form that for a conventional elasticity
problem would lead to a symmetric formulation. Symmetry is however not of
relevance here as the functions $q^{h}$ and $\eps_\mathrm{eq}^{h}$ will
generally come from different function spaces. This formulation is general for
the case in which the space $\mathcal{W}^{h}$ contains discontinuous functions.
However, note if all functions in the space $\mathcal{W}^{h}$ are $C^{0}$
continuous, the formulation is still valid, with terms relating to the jump in
$\eps_{\rm eq}^{h}$ remaining. The formulation would then resemble a
continuous/discontinuous Galerkin method \citep{Engeletal:2002}.

    The solution of the gradient enhanced damage problem requires the
simultaneous solution of equations~\eqref{eqn:equil_cont} and
\eqref{eqn:eps_weak}, which are coupled. Considering the higher-order boundary
condition $\brac{\gradn \brac{ \gradn \eps_{\rm eq}}  \vect{n}} \cdot \vect{n}
=0$ on $\Gamma \cap \Gamma_{d}$, the problem involves: find
$\vect{u}^{h} \in \brac{\mathcal{S}^{h}}^{n}$ and $\Bar{\eps}^{h} \in
\mathcal{W}^{h}$ such that
    \begin{multline}
        \int_{\Omega} \grad \vect{w}^{h} \ip \brac{1 - D \brac{\Bar{\eps}^{h}}}
                \mathbb{C} \ip \grad \vect{u}^{h} \dif \Omega
         - \int_{\Gamma_{h}} \vect{w}^{h} \cdot \vect{h} \dif \Gamma
         - \int_{\Omega} \vect{w}^{h} \cdot \vect{f} \dif \Omega
         \\
         + \int_{\Gamma \cap \Gamma_{d}} {\alpha_{2}} h^{2} \brac{\gradn \cdot
            \brac{\gradn \brac{\gradn \vect{w}^{h}}_{\rm eq} }} \vect{n} \cdot
                E c^{2} \brac{ \gradn \cdot \brac{\gradn
                \eps^{h}_{\rm eq}}} \vect{n} \dif \Gamma
         \\
         = 0
           \quad \forall \ \vect{w}^{h} \in \brac{\mathcal{V}^{h}}^{n} ,
            \label{eqn:equil_cont_b}
    \end{multline}
    \begin{multline}
        \int_{\Omega} q^{h} \Bar{\eps}^{h} \dif \Omega
            -\int_{\Omega} q^{h} \eps_{\mathrm{eq}}^{h} \dif \Omega
            +\int_{\Tilde{\Omega}} \gradn q^{h} \cdot  c^{2}\gradn
            \eps_{\mathrm{eq}}^{h} \dif \Omega
            -\int_{\Gamma} q^{h} c^2 \gradn \eps^{h}_{\rm eq} \cdot \vect{n}
                                            \dif \Gamma \\
            -\int_{\Tilde{\Gamma}} \disc{q^{h}} \cdot
                    c^{2} \ave{\gradn \eps_{\mathrm{eq}}^{h}} \dif \Gamma
            -\int_{\Tilde{\Gamma}} \ave{\gradn q^{h}} \cdot
                        c^{2} \disc{\eps_{\mathrm{eq}}^{h}} \dif \Gamma
                        \\
            + \int_{\Tilde{\Gamma}} \frac{\alpha c^{2}}{h}
            \disc{q^{h}} \cdot {\disc{\eps_{\mathrm{eq}}^{h}}} \dif \Gamma
            =
            0
            \quad \forall \ q^{h} \in \mathcal{W}^{h},
                 \label{eqn:eps_weak_b}
    \end{multline}
where $\alpha_{2}$ is a penalty parameter which attempts to impose the boundary
condition $\brac{\gradn \brac{ \gradn \eps_{\rm eq}} \vect{n}} \cdot \vect{n} =
0$, and $E$ is Young's modulus. The enforcement of this boundary condition is
not in the spirit of
discontinuous Galerkin methods, as the resulting Galerkin problem is not
consistent, in the same sense that penalty methods are not
consistent (the restoration of consistency would require an approach analogous
to Nitsche's method).
However, this is of no consequence for the considered problems due to a
fortuitous choice of interpolation, as will be shown later. Preserving
consistency while imposing the non-standard boundary condition weakly is complex
due to the nonlinear dependency on~$\Bar{\eps}$. This is a topic requiring
further investigation. Boundary conditions are not explicitly applied on
$\Gamma_{d}-\Gamma$ as the necessary conditions are implied implicitly in the
formulation, as will be shown in examining consistency of the proposed
formulation. The weak enforcement of the non-standard boundary condition
proposed here has not been tested numerically.

The nonlinear equations in \eqref{eqn:equil_cont_b} and \eqref{eqn:eps_weak_b}
are coupled through the dependence of $D$ on $\Bar{\eps}^{h}$ and the dependency
of $\Bar{\eps}^{h}$ on~$\vect{u}^{h}$. Linearisation of these equations is
discussed in \citet{Wellsetal:2003}.

\subsection{Consistency of the discontinuous formulation}
    Having added non-standard terms to the weak form, it is important to prove
consistency of the method. Applying integration by parts to the integral over
$\Tilde{\Omega}$ in equation (\ref{eqn:eps_weak}) yields
    \begin{multline}
        \int_{\Tilde{\Omega}} \gradn q^{h} \cdot c^{2} \gradn
                    \eps_\mathrm{eq}^{h} \dif \Omega
            =
            -\int_{\Tilde{\Omega}} q^{h} c^{2} \gradn^{2}
                                \eps_{\mathrm{eq}}^{h}
                    \dif \Omega
            +\int_{\Gamma} q^{h} c^{2} \gradn \eps^{h} _{\mathrm{eq}} \cdot
                        \vect{n} \dif \Gamma  \\
            +\int_{\Tilde{\Gamma}} \ave{q^{h}}
                     c^{2} \disc{\gradn \eps^{h}_{\mathrm{eq}}} \dif \Gamma
            +\int_{\Tilde{\Gamma}} \disc{q^{h}} \cdot
                     c^{2} \ave{\gradn \eps^{h}_{\mathrm{eq}}} \dif \Gamma.
    \end{multline}
Inserting this expression into equation~\eqref{eqn:eps_weak}, and employing
standard variational arguments, the following Euler-Lagrange equations can be
identified:
    \begin{align}
        \Bar{\eps} - \eps_\mathrm{eq} - c^{2} \gradn^{2} \eps_{\mathrm{eq}} &= 0
                                && {\rm in} \ \Tilde{\Omega},
                \label{eqn:EL_a}\\
         c^{2} \disc{\eps_{\mathrm{eq}}} \cdot \vect{n}         &= 0
                                && {\rm on}  \ \Tilde{\Gamma},
                \label{eqn:EL_b} \\
         c^{2} \disc{\gradn \eps_{\mathrm{eq}}}                     &= 0
                                && {\rm on}  \ \Tilde{\Gamma}.
                \label{eqn:EL_c}
    \end{align}
Equation~\eqref{eqn:EL_a} is the original problem over element interiors (see
equation~\eqref{epsbar}). Equations~\eqref{eqn:EL_b} and~\eqref{eqn:EL_c}
impose continuity of the corresponding fields across element boundaries.
The Galerkin form (equation \eqref{eqn:eps_weak}) can therefore be seen as the
weak imposition of these Euler-Lagrange equations.

%---------------------------------------------------------------------------

\section{Analysis and examples}
\label{sect4}

    The formulation outlined in Section~\ref{sect3} has been analysed, and tested
for various problems. Elements are differentiated on the basis of the
interpolation order for $\vect{u}^{h}$ (which is always $C^{0}$ continuous), the
interpolation
order for $\Bar{\eps}^{h}$, and the continuity of $\Bar{\eps}^{h}$. For
example, an element with cubic shape functions for $\vect{u}^{h}$ and $C^{0}$
continuous, quadratic shape functions for $\Bar{\eps}^{h}$ is denoted
$P^3/P^2(C^{0})$. An element with quadratic shape functions for $\vect{u}^{h}$
and discontinuous, linear shape functions for $\Bar{\eps}^{h}$ is  denoted
$P^2/P^1(C^{-1})$.

For one-dimensional examples,  $\eps_\mathrm{eq} =\eps$.

%----------------------------------------

\subsection{Convergence for the elastic case in one dimension}
    For an elastic problem, there exists a one-way coupling between the two
weak equations~\eqref{eqn:equil_cont_b} and~\eqref{eqn:eps_weak_b}. In this
section,  the approximation of $\Bar{\eps}$, for given a solution to
equation~\eqref{eqn:equil_cont_b}, is examined. For an elastic bar, the
fundamental problem is a second-order differential equation, and the second weak
equation  provides a projection of the solution to the equilibrium equation (and
its relevant derivatives) onto the basis for~$\Bar{\eps}$.

\subsubsection{Error analysis for the mixed strain field}
    Here, an analysis for the $L^2$-error in $\Bar{\eps}^{h}$, $\| \Bar{\eps}^{h} -
\eps - c^{2} \eps_{,xx}\|$, where $\eps$ is the exact
strain field, is performed. In order to compare with numerical solutions, we
have solved an elastic problem with a forcing function chosen such that
$\eps$
has a localised character to it. The analysis rests upon the crucial result that
the one-dimensional finite element displacement field, $u^h$, is nodally-exact
when the shape functions are derived from Lagrange polynomials
\citep{strangfix}.

Consider the interpolation combinations $P^{k+1} / P^k(C^0)$ and $P^{k+1} / P^k
(C^{-1})$. Let the nodal values of $\Bar{\eps}^{h}$ corresponding to element $e$
be denoted by
$\delta_{k(e-1)+1}$,  $\dots$, $\delta_{ke+1}$
for the $C^0$ case, and by
$\delta_{(k+1)(e-1)+1}$, $\dots$, $\delta_{(k+1)e}$
for the $C^{-1}$ case. The values of $u^h$ at the nodes corresponding to this
element are
$d_{(k+1)(e-1)+1}$, $\dots$, $d_{(k+1)e+1}$.
The nodal displacements coincide with the exact solution. The discrete matrix
form of equation~\eqref{eqn:eps_weak_b} reveals the stencil relating
the $\delta$- and $d$-values. In the $C^0$ case,
$\{ \delta_{k(e-1)+1}, \dots, \delta_{ke+1} \}$
depend on
$\{ d_{(k+1)(e-3)+1} $, $\dots$, $ d_{(k+1)(e+2)+1} \}$,
and in the $C^{-1}$ case,
$\{ \delta_{(k+1)(e-1)+1}$, $\dots$, $\delta_{(k+1)e} \}$
depend on
$\{ d_{(k+1)(e-3)+1}$, $\dots$, $d_{(k+1)(e+2)+1}\}$;
i.e., on $5k+6$ $d$-values.

\begin{figure}
    \centering
    \psfrag{H}{\tiny$h$}
    \psfrag{A}{\tiny Inter-element $d$-node}
    \psfrag{B}{\tiny Mid-element $d$-node}
    \psfrag{C}{\tiny Inter-element $\delta$-node}
    \psfrag{U}{\tiny $u$}
    \psfrag{V}{\tiny$u^{\rm p}$}
    \psfrag{D1}{\tiny $d_{2e-5}$}
    \psfrag{G}{\tiny$d_{2e+5}$}
    \psfrag{E}{\tiny$\delta_{e-2}$}
    \psfrag{F}{ \tiny$\delta_{e+3}$}
    \includegraphics[width=7cm]{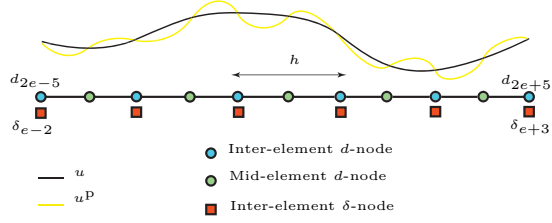}
    \caption{Element patch for error analysis of $P^2/P^1(C^0)$ formulation.}
    \label{figc0}
\end{figure}

\begin{figure}
    \centering
    \psfrag{H}{\tiny$h$}
    \psfrag{A}{\tiny Inter-element $d$-node}
    \psfrag{B}{\tiny Mid-element $d$-node}
    \psfrag{C}{\tiny Inter-element $\delta$-node}
    \psfrag{U}{\tiny$u$}
    \psfrag{V}{\tiny$u^{\rm p}$}
    \psfrag{D1}{\tiny$d_{2e-5}$}\psfrag{G}{\tiny$d_{2 e+5}$}
    \psfrag{E}{\tiny$\delta_{2e- 5}$}
    \psfrag{F}{\tiny$\delta_{2e+4}$}
    \includegraphics[width=7cm]{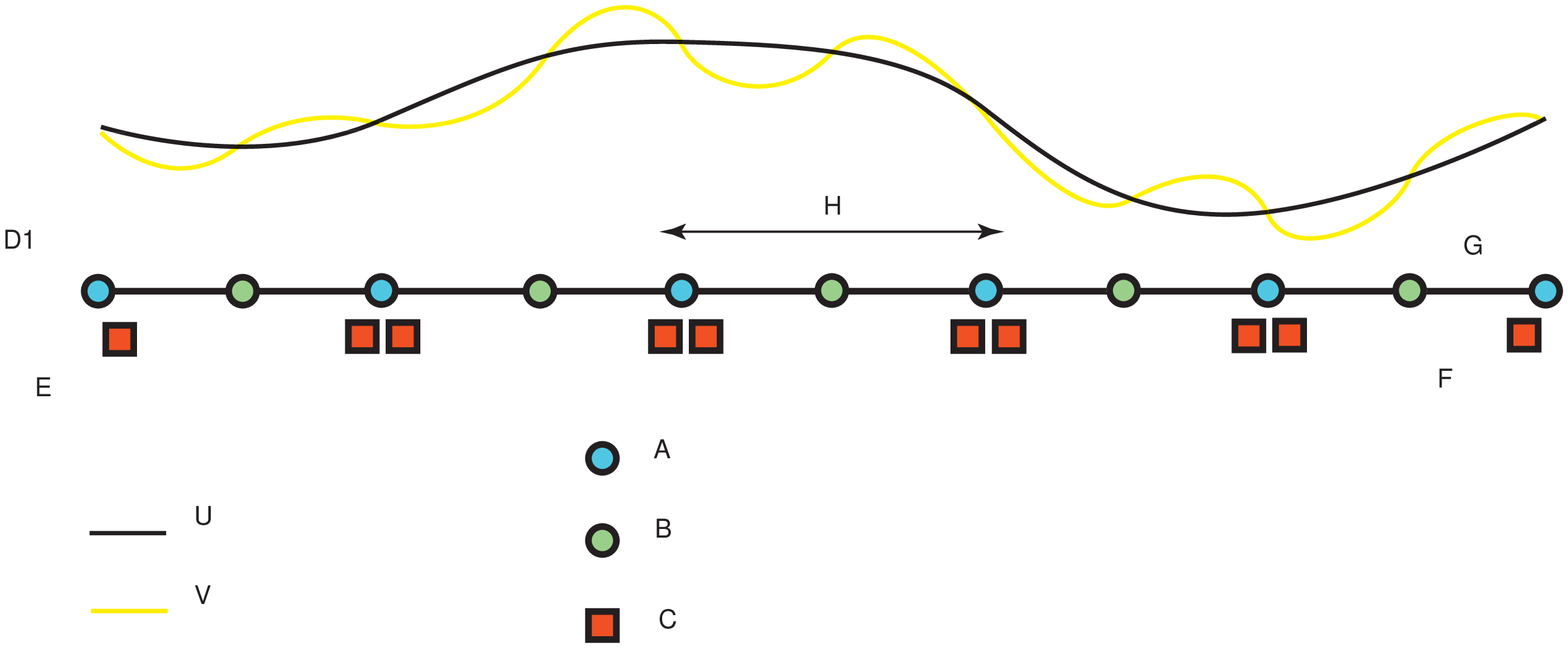}
    \caption{Element patch for error analysis of $P^2/P^1(C^{-1})$ formulation.}
    \label{figc-1}
\end{figure}

Consider a nodally-exact, $5(k+1)^\mathrm{th}$-order polynomial interpolant of
the exact solution. This field is denoted by $u^\mathrm{p}$. Figure~\ref{figc0}
shows a patch of 5 elements, nodes for the $d$- and $\delta$- degrees of
freedom, $u^\mathrm{ex}$ and $u^\mathrm{p}$ for the $P^2/P^1(C^0)$ interpolation
combination. Figure~\ref{figc-1} depicts the situation for the $P^2/P^1(C^{-1})$
interpolation combination.

Let $\eps^{\rm p}$ denote the strain field arising from
$u^{\rm p}$. Using the triangle inequality the error can be
bounded as follows:
\begin{equation}
    \| \Bar{\eps}^{h} - \eps - c^2\eps_{,xx} \|_{\Omega}
    \le \| \Bar{\eps}^{h} - \eps^{\rm p} - c^2\eps^{\rm p}_{,xx} \|_{\Omega}
    + \| \eps^{\rm p} + c^2\eps^{\rm p}_{,xx} - \eps
        - c^2\eps_{,xx} \|_{\Omega}.
    \label{err1}
\end{equation}
Since $u^{\rm p}$ is of order $5(k+1)$, and is nodally-exact over $\Omega_e$,
finite element interpolation theory \citep{odencarey4} leads to the result
\begin{equation}
    \| \eps^{\rm p} - c^2\eps^{\rm p}_{,xx} - \eps -
    c^2\eps_{,xx} \|_{\Omega} \le C h^{5(k+1) - 2},
    \label{err2}
\end{equation}
where $C$ is a constant, independent of $h$.

The first term on the right-hand side in~\eqref{err1} is estimated using the
stencil for each interpolation combination. The $5(k+1)^{\rm th}$-order
polynomial interpolant, $u^{\rm p}(x)$, can be written as
\begin{equation}
    u^{\rm p}(x) = \sum \limits_{n = 0}^{5(k+1)} A_n \brac{x -
                        (e + \frac{1}{2}) h}^n,
    \label{err3}
\end{equation}
where $(e+1/2)h$ is the midpoint of $\Omega_e$ with the first node of the mesh
at $x = 0$. The nodal exactness of $u^h$ allows
$\{ d_{k(e-3)+1}$, $\dots$,$d_{k(e+2)+1}\}$
to be written in terms of~$A_n$.
On combining with the stencil,
$\{\delta_{k(e-1)+1}$, $\dots$, $\delta_{ke+1}\}$
(for the $C^0$ case) and
$\{\delta_{(k+1)(e-1)+1}$, $\dots$, $\delta_{(k+1)e}\}$
(for the $C^{-1}$ case) can therefore be expressed in terms of~$A_n$. Using the
interpolation functions for $\Bar{\eps}^h$, it is then
a trivial matter to exactly evaluate
$\| \Bar{\eps}^{h} - \eps^{\rm p} - c^2\eps^{\rm p}_{,xx} \|_{\Omega_e}$.

For convenience, $\| \Bar{\eps}^{h} - \eps^{\rm p} - c^2\eps^{\rm p}_{,xx}
\|^2_{\Omega_e}$ was evaluated for each interpolation combination. The results
are:
\begin{itemize}
\item {$P^{3}/P^{2}(C^0)$}
    \begin{equation}
    \| \Bar{\eps}^{h} - \eps^{\rm p} - c^2\eps^{\rm p}_{,xx} \|^2_{\Omega_e}
    =
    \brac{\frac{9}{500} {A_3}^2 + 2 A_3 A_5 c^2 + \frac{500}{9} {A_5}^2 c^4} h^5
        + O(h^7);
    \label{errp3p2c0}
    \end{equation}
\item {$P^{2}/P^{1}(C^0)$}
    \begin{equation}
    \| \Bar{\eps}^{h} - \eps^{\rm p} - c^2\eps^{\rm p}_{,xx} \|^2_{\Omega_e}
    =
    \brac{\frac{21}{20} {A_3}^2 + 22 A_3 A_5 c^2 + 120 {A_5}^2 c^4} h^5
        + O(h^7);
    \label{errp2p1c0}
    \end{equation}
\item {$P^{3}/P^{2}(C^{-1})$}
    \begin{equation}
    \| \Bar{\eps}^{h} - \eps^{\rm p} - c^2\eps^{\rm p}_{,xx}\|^2_{\Omega_e}
    =
    \brac{\frac{1936}{27} - \frac{704}{27}\alpha + \frac{64}{27}\alpha^2} {A_4}^2
            c^4 h^3 + O(h^5);
    \label{errp3p2c-1}
    \end{equation}
\item {$P^{2}/P^{1}(C^{-1})$}
    \begin{multline}
    \| \Bar{\eps}^{h} - \eps^{\rm p} - c^2\eps^{\rm p}_{,xx} \|^2_{\Omega_e}
    =
    \brac{\frac{4}{27}{A_2}^2 - \frac{4}{3} A_2 A_4 c^2 + 3 {A_4}^2 c^4} h^3 \\
        + \brac{\frac{4}{3} A_2 A_4 c^2 - 6 {A_4}^2 c^4} \alpha h^3 + 3 {A_4}^2 c^4
                \alpha^2 h^3; \ \text{and}
    \label{errp2p1c-1}
    \end{multline}
\item {$P^{1}/P^{0}(C^{-1})$}
    \begin{equation}
    \| \Bar{\eps}^h - \eps^{\rm p} - c^2\eps^{\rm p}_{,xx} \|^2_{\Omega_e}
    =
    36\brac{1 - 2 \alpha + \alpha^2} {A_3}^2 c^4 h + O(h^3).
    \label{errp1p0}
    \end{equation}
\end{itemize}

For each case listed above, the higher-order terms in $h$ have a finite maximum
power, $O(h^l)$, and $h \le m(\Omega)$, where $m(\Omega)$ is a measure of the
length of the total domain. Therefore it follows that there exist
constants $\widetilde{C}_1$, $\widetilde{C}_2$, $\widetilde{C}_3$,
$\widetilde{C}_{31}$, $\widetilde{C}_4$,$ \widetilde{C}_5$,
$\widetilde{C}_{51}$, such that for the $P^{3}/P^{2}(C^{0})$ element
\begin{equation}
    \| \Bar{\eps}^h - \eps^{\rm p} - c^2\eps^{\rm p}_{,xx} \|^2_{\Omega_e}
        \le \widetilde{C}_1 h^5,
\end{equation}
    for the $P^{2}/P^{1}(C^{0})$ element
\begin{equation}
    \| \Bar{\eps}^h - \eps^{\rm p} - c^2\eps^{\rm p}_{,xx} \|^2_{\Omega_e}
    \le \widetilde{C}_2 h^5,
\end{equation}
    for the $P^{3}/P^{2}(C^{-1})$ element
\begin{equation}
    \|\Bar{\eps}^h - \eps^{\rm p} - c^2\eps^{\rm p}_{,xx} \|^2_{\Omega_e}
    \le
    \begin{cases}
        \widetilde{C}_3 h^5         & {\rm if} \ \alpha = 5.5 \\
        \widetilde{C}_{31} h^3      & {\rm otherwise},
    \end{cases}
\end{equation}
    for the $P^{2}/P^{1}(C^{-1})$ element
\begin{equation}
    \| \Bar{\eps}^h - \eps^{\rm p} - c^2\eps^{\rm p}_{,xx} \|^2_{\Omega_e}
    \le \widetilde{C}_4 h^3 \quad \forall \ \alpha,
\end{equation}
    and for the $P^{1}/P^{0}(C^{-1})$ element,
\begin{equation}
    \| \Bar{\eps}^h - \eps^{\rm p} - c^2\eps^{\rm p}_{,xx} \|^2_{\Omega_e}
    \le
    \begin{cases}
        \widetilde{C}_5 h^3         & {\rm if} \ \alpha = 1 \\
        \widetilde{C}_{51} h        & {\rm otherwise},
    \end{cases}
\end{equation}
where the constants $\widetilde{C}_1, \dots, \widetilde{C}_{51}$ are independent
of $h$ (if $u$ is sufficiently regular) and element number~$e$
(since the constants can be chosen to be the maximum over all elements). For
elements of a uniform size, in each mesh there are $n_{\rm el} =
m(\Omega)/h$ elements. Therefore,
\begin{equation}
    \| \Bar{\eps}^h - \eps^{\rm p} - c^2\eps^{\rm p}_{,xx}\|^2_{\Omega}
    \le
    \frac{m(\Omega)}{h}\, \underset{e}{\max} \| \Bar{\eps}^h - \eps^{\rm p} -
        c^2\eps^{\rm p}_{,xx} \|^2_{\Omega_e},
    \label{err5}
\end{equation}
leading to, for the $P^3/P^2(C^0)$ element
\begin{equation}
    \| \Bar{\eps}^h - \eps^{\rm p} - c^2\eps^{\rm p}_{,xx} \|^2_{\Omega}
    \le
    m(\Omega)\widetilde{C}_1 h^4,
    \label{err61}
\end{equation}
for the $P^2/P^1(C^0)$ element
\begin{equation}
    \| \Bar{\eps}^h - \eps^{\rm p} - c^2\eps^{\rm p}_{,xx}\|^2_{\Omega}
    \le m(\Omega)\widetilde{C}_2 h^4,
\end{equation}
for the $P^3/P^2(C^{-1})$ element
\begin{equation}
    \| \Bar{\eps}^h - \eps^{\rm p} - c^2\eps^{\rm p}_{,xx} \|^2_{\Omega}
    \le
    \begin{cases}
        m(\Omega)\widetilde{C}_3 h^4        & {\rm if} \ \alpha = 5.5\\
       m(\Omega)\widetilde{C}_{31} h^2      & {\rm otherwise},
    \end{cases}
\end{equation}
for the $P^2/P^1(C^{-1})$ element
\begin{equation}
    \| \Bar{\eps}^h - \eps^{\rm p} - c^2\eps^{\rm p}_{,xx} \|^2_{\Omega_e}
    \le
    m(\Omega)\widetilde{C}_4 h^2 \quad \forall \ \alpha,
\end{equation}
and for the $P^1/P^0(C^{-1})$ element
\begin{equation}
    \| \Bar{\eps}^h - \eps^{\rm p} - c^2\eps^{\rm p}_{,xx}\|^2_{\Omega_e}
    \le
    \begin{cases}
        m(\Omega)\widetilde{C}_5 h^2        & {\rm if} \ \alpha = 1 \\
        m(\Omega)\widetilde{C}_{51} h^0     & {\rm otherwise}.
    \end{cases}
    \label{err65}
\end{equation}

Since $k \ge 0$ in~\eqref{err2}, each of equations
\eqref{err61}--\eqref{err65} can be combined with \eqref{err2}, and
constants $C_1$, $C_2$, $C_3$, $C_{31}$, $C_4$, $C_5$, $C_{51}$ can be found,
independent of $h$, such that for the $P^3/P^2(C^0)$ element
\begin{equation}
    \| \Bar{\eps}^h - \eps^{\rm p} - c^2\eps^{\rm p}_{,xx}\|^2_{\Omega}
    \le m(\Omega)C_1 h^4,
\end{equation}
for the $P^2/P^1(C^0)$ element
\begin{equation}
    \| \Bar{\eps}^h - \eps^{\rm p} - c^2\eps^{\rm p}_{,xx} \|^2_{\Omega}
    \le m(\Omega)C_2 h^4,
\end{equation}
for the $P^3/P^2(C^{-1})$ element
\begin{equation}
    \| \Bar{\eps}^h - \eps^{\rm p} - c^2\eps^{\rm p}_{,xx}\|^2_{\Omega}
    \le
    \begin{cases}
        m(\Omega)C_3 h^4        & {\rm if} \ \alpha = 5.5 \\
        m(\Omega)C_{31} h^2     & {\rm otherwise},
    \end{cases}
\end{equation}
for the $P^2/P^1(C^{-1})$ element
\begin{equation}
    \| \Bar{\eps}^h - \eps^{\rm p} - c^2\eps^{\rm p}_{,xx} \|^2_{\Omega}
    \le m(\Omega)C_4 h^2 \quad \forall \ \alpha,
\end{equation}
and for the $P^1/P^0(C^{-1})$ element,
\begin{equation}
    \| \Bar{\eps}^h - \eps^{\rm p} - c^2\eps^{\rm p}_{,xx}\|^2_{\Omega}
    \le
    \begin{cases}
        m(\Omega)C_5 h^2        & {\rm if} \ \alpha = 1 \\
        m(\Omega)C_{51} h^0     & {\rm otherwise}.
    \end{cases}
\label{err7}
\end{equation}
The results of the analysis are summarised in Table~\ref{tab:conv}.
    \begin{table}
        \begin{center}
        \begin{tabular}{|l|l|}
            \hline
            Element type                            & convergence rate \\ \hline
            $P^{1}/P^{0}(C^{-1})$ ($\alpha=1$)      & $O\brac{h^{1}}$ \\
            $P^{1}/P^{0}(C^{-1})$ ($\alpha \ne 1$)  & $O\brac{h^{0}}$ \\
            $P^{2}/P^{1}(C^{0})$                    & $O\brac{h^{2}}$ \\
            $P^{2}/P^{1}(C^{-1})$                   & $O\brac{h^{1}}$ \\
            $P^{3}/P^{2}(C^{0})$                    & $O\brac{h^{2}}$ \\
            $P^{3}/P^{2}(C^{-1})$ ($\alpha=5.5$)    & $O\brac{h^{2}}$ \\
            $P^{3}/P^{2}(C^{-1})$ ($\alpha \ne 5.5$)& $O\brac{h^{1}}$ \\ \hline
        \end{tabular}
        \end{center}
        \caption{Analytical convergence rates in terms of
            $\| \Bar{\eps} -\Bar{\eps}^{h} \|_{\Omega}$
            for various elements.}
    \label{tab:conv}
    \end{table}

\subsubsection{Observed convergence rates}
\begin{figure}
        \centering
        \includegraphics[width=0.9\textwidth]{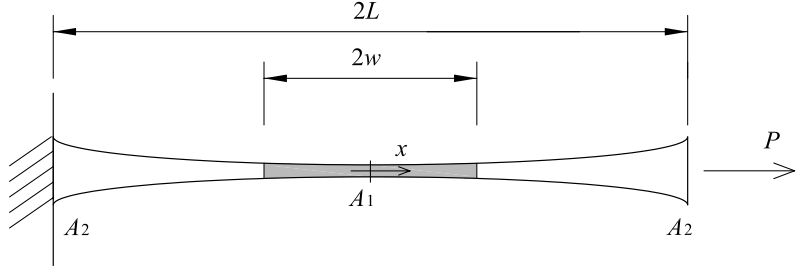}
    \caption{Quadratically tapering bar.}
    \label{fig:tapered_bar}
    \end{figure}

    The convergence rate estimates from the previous section are now compared with
observed rates. To do this, the convergence of $\Bar{\eps}$ is examined for the
quadratically tapering bar shown in Figure~\ref{fig:tapered_bar}.
Considering that the origin is located as the centre of the bar, the cross
section $A$ is given by
    \begin{equation}
        A(x) = A_{1} + \gamma^{2} A_{1} \frac{x^{2}}{L^{2}},
    \end{equation}
where $\gamma$ controls the ratio between $A_{1}$ and $A_{2}$ ($\gamma^{2} =
\brac{A_{2}-A_{1}} /A_{1}$).
This problem can be equivalently formulated as a rod with unit cross-section
and loading $f$ which varies along the rod, which allows the convergence
analysis from the previous section to be carried over for this case.
The exact solution for $\eps$ along the bar is:
    \begin{equation}
        \eps = \frac{P L^{2} }{ EA_{1}\brac{L^{2} + \gamma^{2} x^{2}}},
    \end{equation}
and the exact solution of $\eps_{,xx}$ is:
    \begin{equation}
        \eps_{,xx}
        =
        \frac{2 P L^{2} \gamma^{2} \brac{3\gamma^{2}x^{2}-L^{2}}}
        {EA_{1}\brac{L^{2} + \gamma^{2} x^{2}}^{3}}.
    \end{equation}

    The term $\Bar{\eps}^{h}$ involves
both $\eps$ and $\eps_{,xx}$.  The $\eps$ component
involves the standard $L^{2}$ projection of $\eps^{h}$ (coming from the
solution of the equilibrium equation) onto the basis for $\Bar{\eps}^{h}$.
This projection does not involve any of the element interface terms that have
arisen in the discontinuous Galerkin formulation. Of special interest is the
approximation of the $\eps_{,xx}$ term. To examine numerically the convergence
of this term, the weak form corresponding to:
    \begin{equation}
        \Bar{\eps} - \eps_{,xx} = 0
    \end{equation}
is solved, as it is the $\eps_{,xx}$ term that dominates the
convergence rate. However, in practice,
the convergence rate of $\eps$ may appear to dominate due to a large
difference in the constants in the error inequality. Taking
equation~\eqref{eqn:eps_weak_b} and removing terms related to
the projection of $\eps^{h}$, we solve the following problem: find $u^{h}
\in \mathcal{S}^{h}$ and $\Bar{\eps}^{h} \in \mathcal{W}^{h}$ such that
    \begin{equation}
        \int_{\Omega} w^{h}_{,x} E u^{h}_{,x} \dif \Omega
                                                - w^{h} P |_{x=L} = 0
            \quad \forall \ w^{h} \in \mathcal{V}^{h},
    \end{equation}
    \begin{multline}
        \int_{\Omega} q^{h} \Bar{\eps}^{h} \dif \Omega
            +\int_{\Tilde{\Omega}} q^{h}_{,x} \eps^{h}_{,x} \dif \Omega
            -\int_{\Gamma} q^{h} \eps_{,x}^{h} n \dif \Gamma
            -\int_{\Tilde{\Gamma}} \disc{q^{h}}
                    \ave{\eps^{h}_{,x}} \dif \Gamma \\
            -\int_{\Tilde{\Gamma}} \ave{q^{h}_{,x}}
                        \disc{\eps^{h}} \dif \Gamma
         + \int_{\Tilde{\Gamma}} \frac{\alpha}{h}
            \disc{q^{h}} \disc{\eps^{h}} \dif \Gamma
            = 0 \quad \forall \ q^{h} \in \mathcal{W}^{h}.
    \end{multline}

\begin{figure}
        \psfrag{x}{$x$}
        \psfrag{y}{$\eps_{,xx}$}
        \center \includegraphics{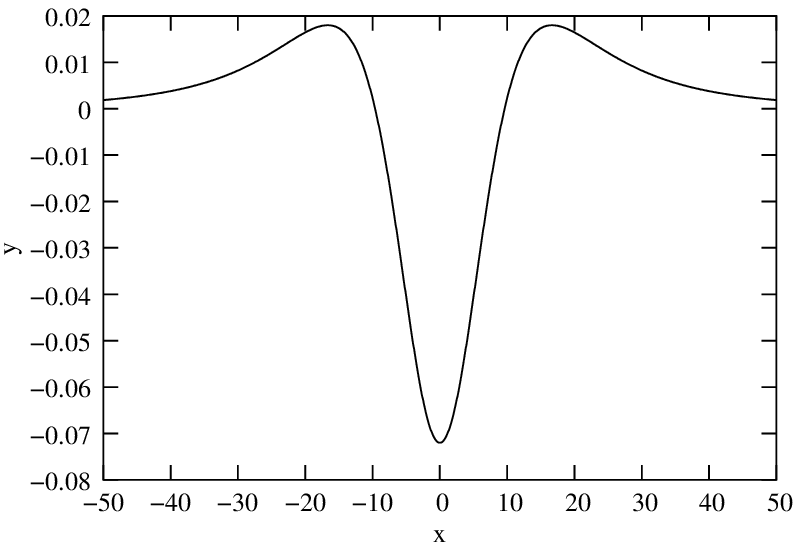}
    \caption{Exact solution for $\eps_{,xx}$.}
    \label{fig:eps_xx}
    \end{figure}

\begin{figure}
        \psfrag{x}{$n_{\rm el}$}
        \psfrag{y}{$\|\eps_{,xx} - \Bar{\eps}^{h}\|$}
        \psfrag{P10d}{{\tiny $P^{1}/P^{0}(C^{-1})$}}
        \psfrag{P21c}{{\tiny $P^{2}/P^{1}(C^{0})$}}
        \psfrag{P21d}{{\tiny $P^{2}/P^{1}(C^{-1})$}}
        \psfrag{P32c}{{\tiny $P^{3}/P^{2}(C^{0})$}}
        \psfrag{P32d}{{\tiny $P^{3}/P^{2}(C^{-1})$}}
        \center\includegraphics{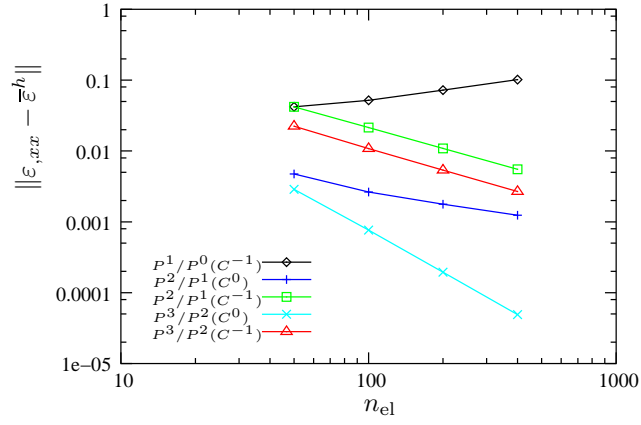}
    \caption{Convergence for the problem $\Bar{\eps} = \eps_{,xx}$ with $\alpha =
                1$ on the domain $-50<x<50$.}
    \label{fig:error_eps_xx}
    \end{figure}

\begin{figure}
        \psfrag{x}{$n_{\rm el}$}
        \psfrag{y}{$\|\eps_{,xx} - \Bar{\eps}^{h}\|$}
        \psfrag{P10d}{{\tiny $P^{1}/P^{0}(C^{-1})$}}
        \psfrag{P21c}{{\tiny $P^{2}/P^{1}(C^{0})$}}
        \psfrag{P21d}{{\tiny $P^{2}/P^{1}(C^{-1})$}}
        \psfrag{P32c}{{\tiny $P^{3}/P^{2}(C^{0})$}}
        \psfrag{P32d}{{\tiny $P^{3}/P^{2}(C^{-1})$}}
        \center\includegraphics{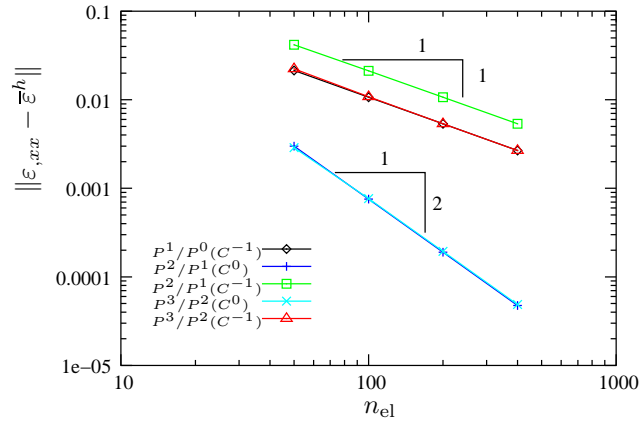}
    \caption{Convergence for the problem $\Bar{\eps} = \eps_{,xx}$ with $\alpha =
            1$ on the domain $-40 < x < 40$.}
    \label{fig:error_eps_xx_b}
    \end{figure}

    For the parameters $A_{1} = 1$~mm$^{2}$, $A_{2} = 0.1$~mm$^{2}$, $L=50$~mm,
$E=1$~MPa and $P=1$~N, the convergence behaviour is examined for a range of
elements. To gain insight, the form of the exact solution is shown in
Figure~\ref{fig:eps_xx}.
The $L^{2}$-norm of the computed error for a range elements is shown in
Figure~\ref{fig:error_eps_xx} for the case~$\alpha = 1$.
If the error is computed by integrating over the entire bar ($-50<x<50$), as in
Figure~\ref{fig:error_eps_xx}, the results are polluted by errors at the
boundaries of the bar. In
Figure~\ref{fig:error_eps_xx}, the convergence behaviour represents primarily
how rapidly the computed solution approaches the exact
solution at the boundaries. However, when simulating a
tapered damaging bar, the error at the boundaries of the bar is of little
consequence as damage develops at the centre, and $\Bar{\eps}$ at the boundaries
plays no role (presuming that the deviation from the exact solution is not
sufficient to induce spurious damage development). Excluding the error at the
end of the
bar by integrating over $-40 < x < 40$, the convergence behaviour is more
predictable, as can be seen in Figure~\ref{fig:error_eps_xx_b}.
From Figure~\ref{fig:error_eps_xx_b}, it can be concluded that the convergence
rate for all elements is consistent with the predicted rates (as summarised in
Table~\ref{tab:conv}).

\begin{figure}
        \psfrag{x}{$n_{\rm el}$}
        \psfrag{y}{$\|\eps_{,xx} - \Bar{\eps}^{h}\|$}
        \psfrag{aaaa1}{{\tiny $\alpha=1$}}
        \psfrag{aaaa2}{{\tiny $\alpha=2$}}
        \psfrag{aaaa3}{{\tiny $\alpha=3$}}
        \psfrag{aaaa4}{{\tiny $\alpha=4$}}
        \psfrag{aaaa5}{{\tiny $\alpha=5$}}
        \center \includegraphics{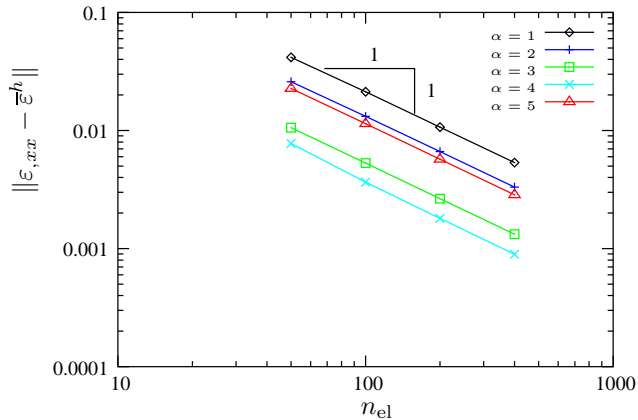}
        \caption{Convergence of the $P^{2}/P^{1}\brac{C^{-1}}$ element for different
            values of $\alpha$ on the domain $-40 < x < 40$ for the problem $\Bar{\eps} =
            \eps_{,xx}$.}
    \label{fig:error_eps_xx_12_alpha}
    \end{figure}

\begin{figure}
        \psfrag{x}{$n_{\rm el}$}
        \psfrag{y}{$\|\eps_{,xx} - \Bar{\eps}^{h}\|$}
        \psfrag{aaaa1}{{\tiny $\alpha=1$}}
        \psfrag{aaaa2}{{\tiny $\alpha=2$}}
        \psfrag{aaaa4}{{\tiny $\alpha=4$}}
        \psfrag{aaa55}{{\tiny $\alpha=5.5$}}
        \psfrag{aaa58}{{\tiny $\alpha=5.8$}}
        \psfrag{aaaa6}{{\tiny $\alpha=6$}}
        \psfrag{aaa62}{{\tiny $\alpha=6.2$}}
        \psfrag{aaa65}{{\tiny $\alpha=6.5$}}
        \center \includegraphics{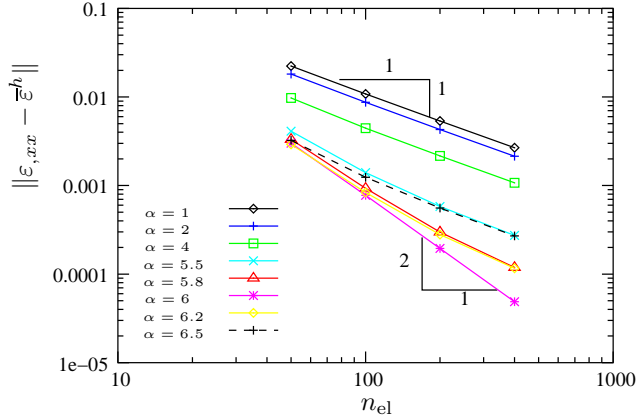}
    \caption{Convergence of the $P^{3}/P^{2}\brac{C^{-1}}$ element for different
            values of $\alpha$ on the domain $-40 < x < 40$ for the problem $\Bar{\eps} =
            \eps_{,xx}$.}
    \label{fig:error_eps_xx_23_alpha}
    \end{figure}

    The effect of $\alpha$ on convergence for the $P^{2}/P^{1}\brac{C^{-1}}$
element is shown in Figure~\ref{fig:error_eps_xx_12_alpha}.
As expected, the convergence rate is unaffected by~$\alpha$. For the
$P^{3}/P^{2}\brac{C^{-1}}$ element, the convergence behaviour for
different $\alpha$ is shown in Figure~\ref{fig:error_eps_xx_23_alpha}.
From Figure~\ref{fig:error_eps_xx_23_alpha}, it is clear the convergence rate
increases by one order for $\alpha=6$. This is close to the predicted value
of~$\alpha=5.5$.

\subsection{Convergence for the inelastic case in one dimension}

    The formulation is now examined for the inelastic case. Again, the
quadratically tapering bar is considered, which leads to damage development at
the centre of the bar. For a particular relationship between $D$ and $\kappa$,
it is possible to solve the problem analytically, which provides the basis for
numerical convergence tests.

\subsubsection{Analytical solution}
\label{sec:exact}
% %
    Consider again the bar shown in Figure~\ref{fig:tapered_bar}, where the
shaded region indicates the damaged zone and $x= \pm w$ is the location of the
damage--elastic boundary. For convenience, the force $P$ is expressed as
    \begin{equation}
        P = \brac{1-\beta^{2}} EA_{1}\kappa_{0},
    \label{eqn:load}
    \end{equation}
where $\beta$ governs the magnitude of the applied load and $\kappa_{0}$ is the
value of $\kappa$ at which damage is first induced. In the undamaged part
of the bar, the strain response is given by
    \begin{equation}
        \eps = \dfrac{P}{EA},
    \end{equation}
and within the damaged zone by,
    \begin{equation}
        \eps = \dfrac{P}{\brac{1-D}EA}.
    \label{eqn:dam_strain}
    \end{equation}
With the intention of finding an analytical solution, a simple damage law is
assumed,
    \begin{equation}
        D =
        \begin{cases}
        0                                       & {\rm if} \ \kappa \le \kappa_{0}\\
        1-\dfrac{\kappa_0}{\kappa}              & {\rm if} \ \kappa  > \kappa_{0}.
        \end{cases}
    \label{eqn:simple_dam}
    \end{equation}
The relationship in equation~\eqref{eqn:simple_dam} is the analogy of perfect
plasticity for the case $c=0$, in the sense that it yields a plateau in the
load--displacement response once the elastic limit has been exceeded.
Assuming that no unloading takes place in the damaged zone, $\kappa =
\Bar{\eps}$, that is
    \begin{equation}
        \kappa  = \eps + c^{2} \eps_{,xx} \quad -w < x < w.
    \label{eqn:reg_strain}
    \end{equation}
Inserting equations \eqref{eqn:reg_strain} and~\eqref{eqn:simple_dam} into
equation~\eqref{eqn:dam_strain}, the following ordinary differential
equation is obtained:
    \begin{equation}
        \brac{1 - \dfrac{P}{\kappa_{0}EA}} \eps
            - c^{2} \dfrac{P}{\kappa_{0} EA} \eps_{,xx} = 0
            \quad -w < x < w,
    \end{equation}
which governs the response in the damaged zone.
The general solution to the above equation for $x>0$ is given by:
    \begin{equation}
        \eps = C_{1} \frac{ M \brac{\frac{\beta^{2} L} {4 c \gamma
                \sqrt{1 + \beta^{2}}},
                \dfrac{1}{4},
                \frac{\gamma x^{2}}{ L c \sqrt{1+\beta^{2}} } }}{\sqrt{x}}
            +
               C_{2} \frac{ W \brac{\frac{\beta^{2} L} {4 c \gamma \sqrt{1 +
                \beta^{2}}},
                \dfrac{1}{4},
                \frac{\gamma x^{2}}{L c \sqrt{1+\beta^{2}} } }}{\sqrt{x}},
    \label{eqn:sol1}
    \end{equation}
where $M$ and $W$ are Whittaker's functions~\citep{math_functions:book}, and
$C_{1}$ and $C_{2}$ are integration constants. The integration constants can be
obtained by considering boundary conditions at~$x = \pm w$. The first considered
condition is symmetry about~$x=0$.
Expanding equation~\eqref{eqn:sol1} in a Taylor series about $x=0$, and
requiring that odd terms vanish, the following relation is obtained:
    \begin{equation}
        C_{1} = C_{2} \frac{ 2 \pi }{\Gamma \brac{ \frac{1}{4} -
            \frac{-\beta^{2} L^{2} + c \gamma \sqrt{1+\beta^{2} } } {4c
                \sqrt{1 + \gamma \beta^{2}} } }},
    \end{equation}
where $\Gamma$ is the Gamma function. The second considered condition is:
    \begin{equation}
        \disc{\eps_{,x}} = 0 \quad  {\rm at} \ x = w,
    \end{equation}
which implies
    \begin{equation}
        \eps_{,x} = \brac{\frac{P}{EA}}_{,x}  \quad {\rm at} \ x = w.
    \end{equation}
    Finally, the location of the elastic--damage boundary can be determined by
requiring that
    \begin{equation}
        \kappa = \kappa_{0}             \quad {\rm at} \ x = w.
    \end{equation}
For the sake of brevity, the expressions for $C_{1}$, $C_{2}$ and $w$ have been
omitted.

\subsubsection{Damage convergence results}
    The convergence behaviour of various elements for the outlined test problem is
now examined. For the tests, the following parameters are adopted: $\gamma = 3$,
$\beta = 5/29$, $L=100$~mm, $E = 200 \times 10^{3}$~MPa, $c=1$~mm and $A_{1} =
1$~mm$^{2}$ and $\kappa_{0} = 1 \times 10^{-3}$. In calculating the error, the
Gamma function has been computed numerically.

\begin{figure}
        \psfrag{x}{$n_{\rm el}$}
        \psfrag{y}{$\|\eps - \eps^{h} \|$}
        \psfrag{P10d}{{\tiny $P^{1}/P^{0}(C^{-1})$}}
        \psfrag{P21c}{{\tiny $P^{2}/P^{1}(C^{0})$}}
        \psfrag{P21d}{{\tiny $P^{2}/P^{1}(C^{-1})$}}
        \psfrag{P32c}{{\tiny $P^{3}/P^{2}(C^{0})$}}
        \psfrag{P32d}{{\tiny $P^{3}/P^{2}(C^{-1})$}}
        \center\includegraphics{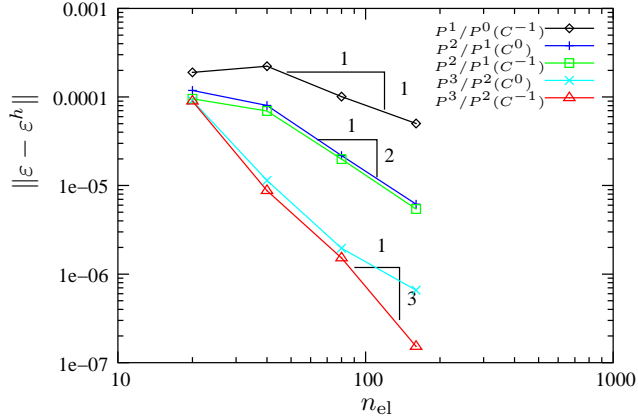}
    \caption{Error in $\eps^{h}$ for the damaging bar.}
    \label{fig:error_eps_dam}
    \end{figure}

The results of the error analysis, for various elements, are shown in
Figure~\ref{fig:error_eps_dam}.
For the $P^{1}/P^{0}\brac{C^{-1}}$, $P^{2}/P^{1}\brac{C^{-1}}$ and
$P^{3}/P^{2}\brac{C^{-1}}$ elements, $\alpha =1$, $\alpha = 4$ and $\alpha=6$,
respectively. These choices draw upon the observed convergence results for the
elastic case. For all the elements, the solution converges. Once a level of
refinement has been reached, the convergence rate is linear.
Moreover, for a given interpolation order, continuous and discontinuous
interpolations yield similar results.

\subsubsection{Non-trivial damage response}
    The quadratically tapering bar is now examined for a non-trivial softening
relationship. For the tapered bar the relevant parameters are: $\gamma= 3$,
$A_{1}=1$~mm$^{2}$,  Young's modulus $E= 20 \times 10^3$~MPa, $\kappa_0=1 \times
10^{-4}$, $\kappa_{c}=0.0125$, and $c=1$~mm. The functional
form of the damage variable is specified to be
\begin{equation}
    D =
    \begin{cases}
        0                                   & {\rm if} \ \kappa \le \kappa_0\\
        1-\dfrac{\kappa_{0}(\kappa_{c}-\kappa)}{\kappa(\kappa_{c}-\kappa_0)}
                                            & {\rm if} \ \kappa_0 < \kappa < \kappa_{c} \\
        1                                   & {\rm if} \ \kappa \ge \kappa_{c},
    \end{cases}
\end{equation}
which leads to a linear softening relationship for $c=0$.
The numerical performance of the formulation has previously been demonstrated
in \citet{Wellsetal:2003} for continuous, piecewise linear $u^h$ and
constant $\Bar{\eps}^h$, discontinuous across element boundaries (the
$P^{1}/P^{0}(C^{-1})$ element). The formulation was shown
numerically to converge to a benchmark solution. Here, the formulation is
extended for a range of different element types.

The motivation behind the considered strain gradient-dependent model is
regularisation in the presence of strain softening. Without strain gradient
effects ($c=0$), computed results are pathologically mesh-dependent; the result
of which is manifest in the load--displacement responses. Therefore, each of the
elements which to this point have been examined are tested, and the
load-displacement responses reported for meshes with 20, 40, 80, 160 and 320
elements. For the $P^3/P^2(C^{-1})$ element, the numerical tests are
performed with meshes of 20, 40, 80 and 160 elements.
To provide a reference solution, the response computed using 160
$P^{3}/P^{2}\brac{C^{0}}$ elements is included in all figures.

\begin{figure}
    \begin{center}
    \begin{tabular}{c}
            \psfrag{x}{$u$}
            \psfrag{y}{$P$}
            \includegraphics{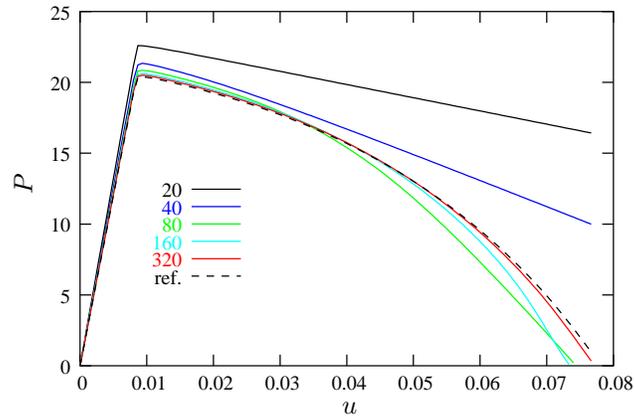} \\ (a) \\
            \psfrag{x}{$u$}
            \psfrag{y}{$P$}
            \includegraphics{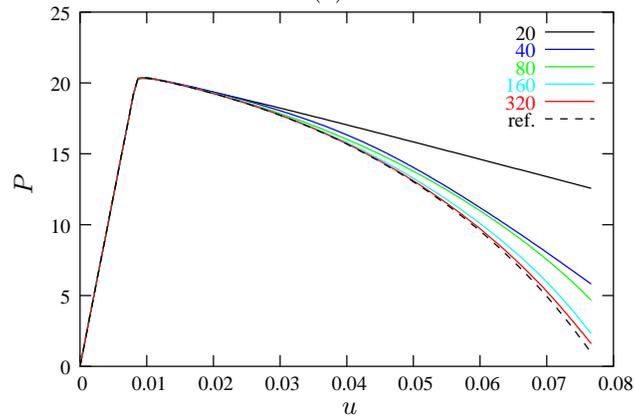} \\ (b)
    \end{tabular}
    \end{center}
    \caption{Load--displacement response for
            (a) $P^{2}/P^{1}\brac{C^{0}}$ and
            (b) $P^{3}/P^{2}\brac{C^{0}}$ elements.}
    \label{fig:Pu_continuous}
\end{figure}

\begin{figure}
    \begin{center}
    \begin{tabular}{c}
            \psfrag{x}{$u$}
            \psfrag{y}{$P$}
            \includegraphics{PU_P1P0.eps} \\ (a) \\
            \psfrag{x}{$u$}
            \psfrag{y}{$P$}
            \includegraphics{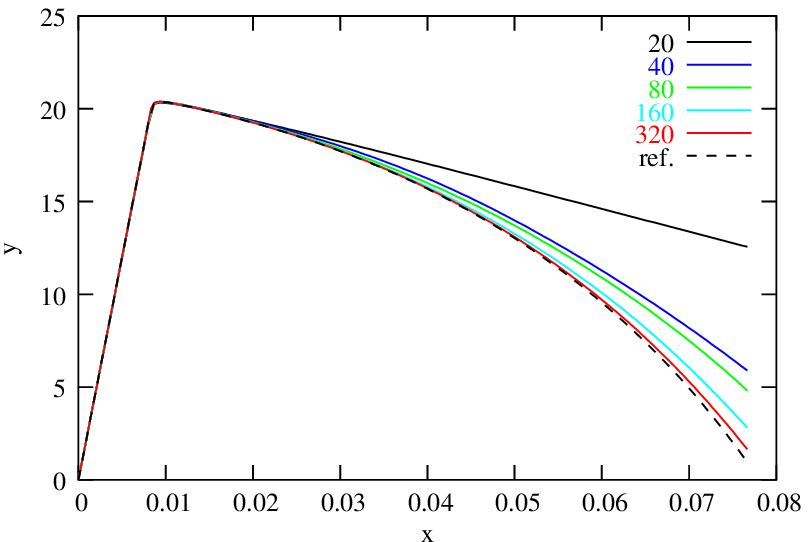} \\[-2ex] (b) \\
            \psfrag{x}{$u$}
            \psfrag{y}{$P$}
            \includegraphics{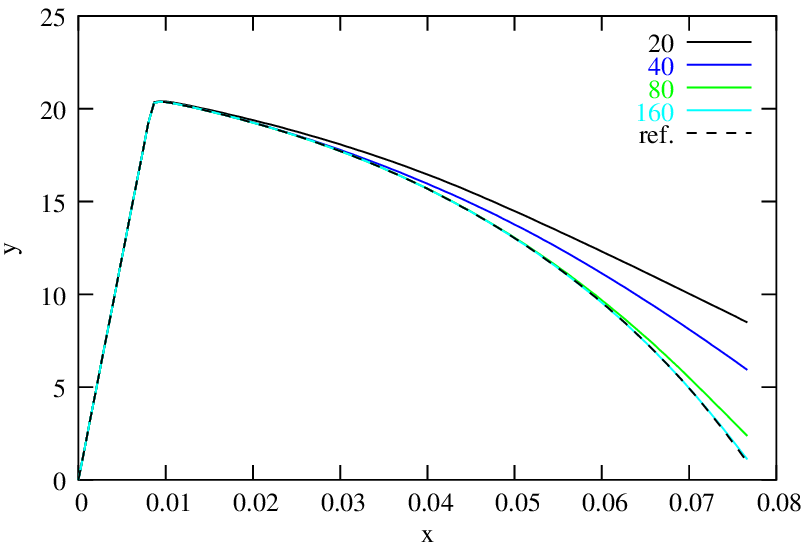} \\[-2ex] (c)
    \end{tabular}
    \end{center}
    \caption{Load--displacement response for
            (a) $P^{1}/P^{0}\brac{C^{-1}}$,
            (b) $P^{2}/P^{1}\brac{C^{-1}}$ and
            (c) $P^{3}/P^{2}\brac{C^{-1}}$ elements.}
    \label{fig:Pu_discontinuous}
\end{figure}

\begin{figure}
        \psfrag{x}{$x$}
        \psfrag{y}{$D$}
        \psfrag{aaaaaP2P1c}{\tiny $P^{2}/P^{1}(C^{0})$}
        \psfrag{aaaaaP3P2c}{\tiny $P^{3}/P^{2}(C^{0})$}
        \center \includegraphics{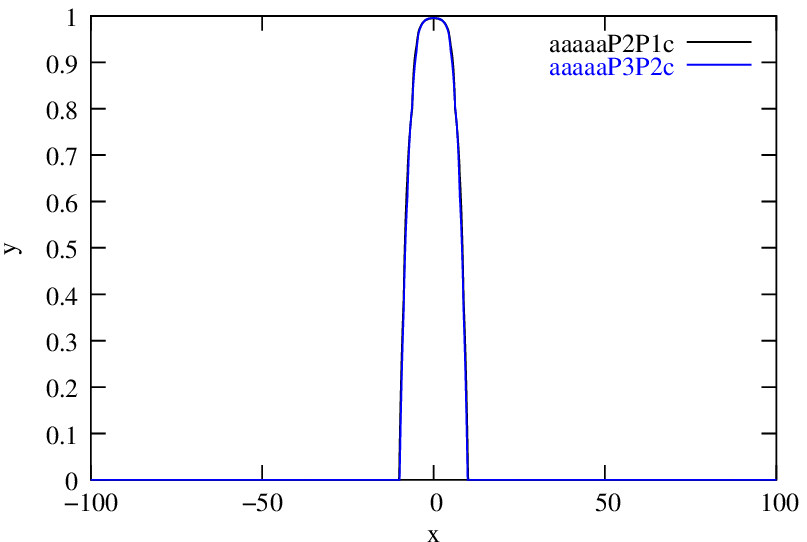}
        \caption{Damage profiles for the $\Bar{\eps}^{h}$-continuous elements,
        computed using 160 elements.}
        \label{fig:D_profile_continuous}
    \end{figure}

\begin{figure}
        \psfrag{x}{$x$}
        \psfrag{y}{$D$}
        \psfrag{aaaaaP1P0d}{\tiny $P^{1}/P^{0}(C^{-1})$}
        \psfrag{aaaaaP2P1d}{\tiny $P^{2}/P^{3}(C^{-1})$}
        \psfrag{aaaaaP3P2d}{\tiny $P^{3}/P^{2}(C^{-1})$}
        \center \includegraphics{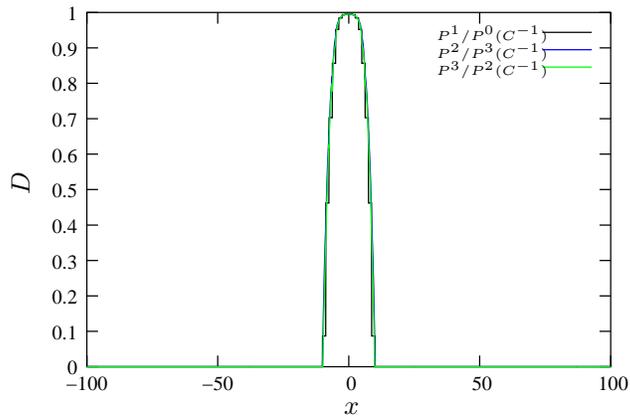}
        \caption {Damage profiles for the $\Bar{\eps}^{h}$-discontinuous
        elements, computed using 160 elements.}
    \label{fig:D_profile_discontinuous}
    \end{figure}

    The load-displacement responses for the two elements using a continuous
interpolation of $\Bar{\eps}^{h}$ are shown in Figure~\ref{fig:Pu_continuous}.
As the mesh is refined, the computed response for both element types converges
towards to the reference solution. The load-displacement responses for three
elements using a discontinuous interpolation of $\Bar{\eps}^{h}$ are shown in
Figure~\ref{fig:Pu_discontinuous}.
Again, for the $P^{1}/P^{0}\brac{C^{-1}}$ element, $\alpha =1$, for the
$P^{2}/P^{1}\brac{C^{-1}}$ element, $\alpha =4$, and for the
$P^{3}/P^{2}\brac{C^{-1}}$ element, $\alpha=6$.
It is clear, for all elements, that the load--displacement response
converges to the reference solution with mesh refinement.
To further examine the computed results, the damage profiles for the two
continuous elements and the three discontinuous elements are compared in
Figures~\ref{fig:D_profile_continuous} and~\ref{fig:D_profile_discontinuous},
respectively. For all cases, the 160 element mesh is considered.
    Clearly, the damage profiles are nearly identical for all element types.

\subsection{Three-point bending test}
    To conclude the numerical validations, a three-point bending test is performed
using a $P^{2}/P^{1}\brac{C^{0}}$ element. The element is triangular, with
degrees of freedom for $\vect{u}^{h}$ located at the vertexes and at the
mid-sides, and degrees of freedom for $\Bar{\eps}^{h}$ located only at the
vertexes of the element. The choice of a quadratic interpolation of
$\vect{u}^{h}$ means
that the penalty term for imposing the non-standard boundary condition in
equation~\eqref{eqn:equil_cont_b} vanishes, and the non-standard boundary
condition is satisfied by construction (see equation~\eqref{eqn:equil_cont_b}).
The equivalent strain is taken as the trace of the strain tensor,
    \begin{equation}
        \eps^{\rm eq} = {\rm trace}\brac{\strain}.
    \end{equation}
This choice does not reflect a strong physical motivation, rather it is chosen
for illustrative purposes as it allows for relatively simple linearisation of
the method (which can become extremely complex when $c \ne 0$).

\begin{figure}
        \center \includegraphics{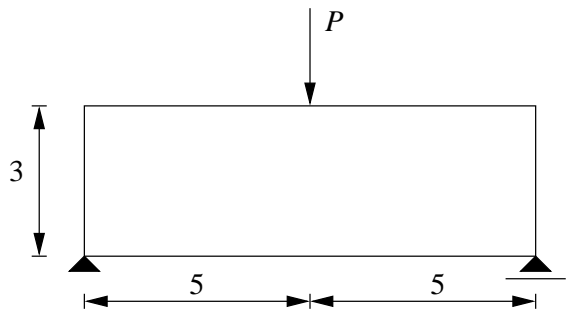}
    \caption{Three-point bending specimen.}
    \label{fig:3bend}
    \end{figure}

\begin{figure}
        \begin{center}
        \begin{tabular}{cc}
            \includegraphics[width=0.47\textwidth]{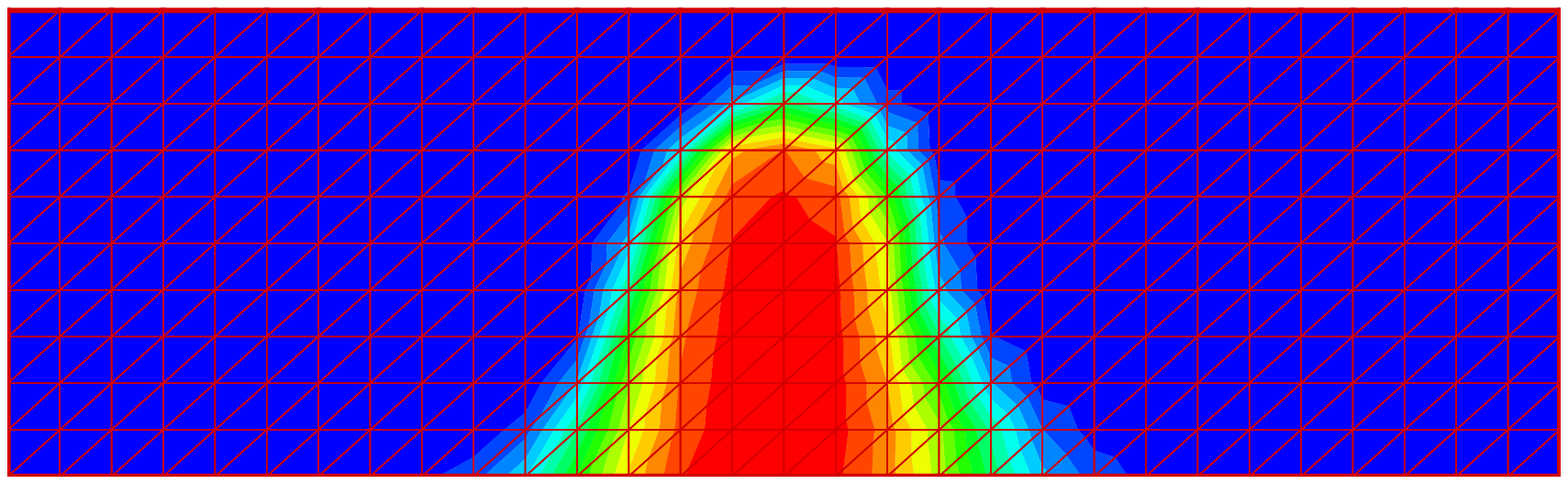} &
            \includegraphics[width=0.47\textwidth]{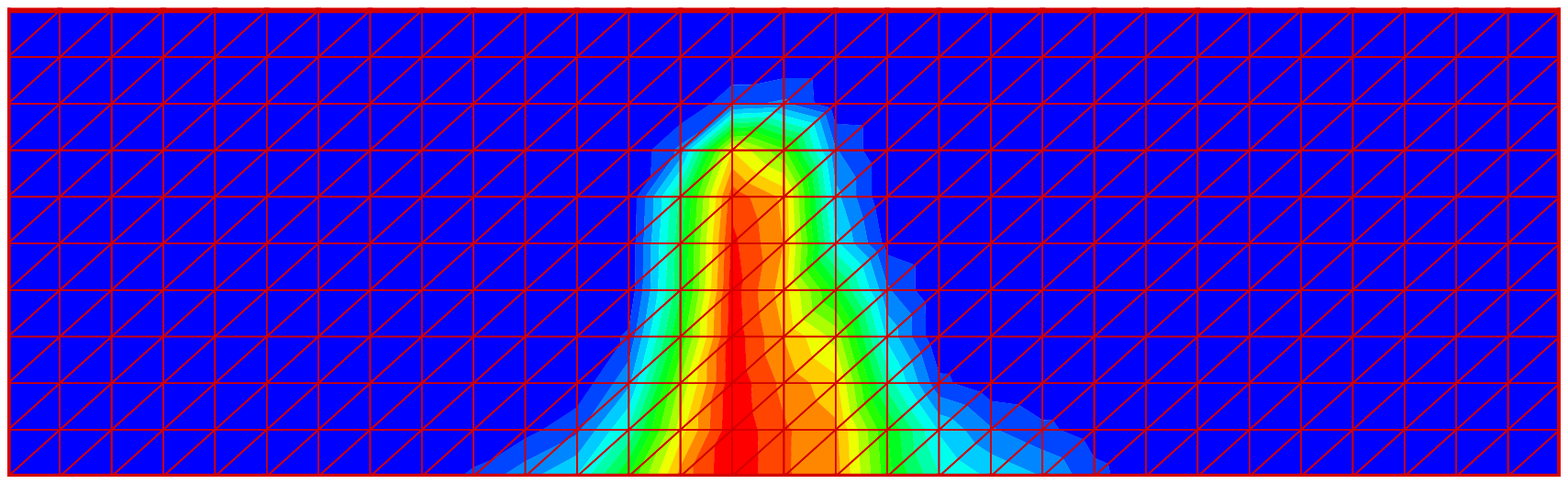} \\
            \includegraphics[width=0.47\textwidth]{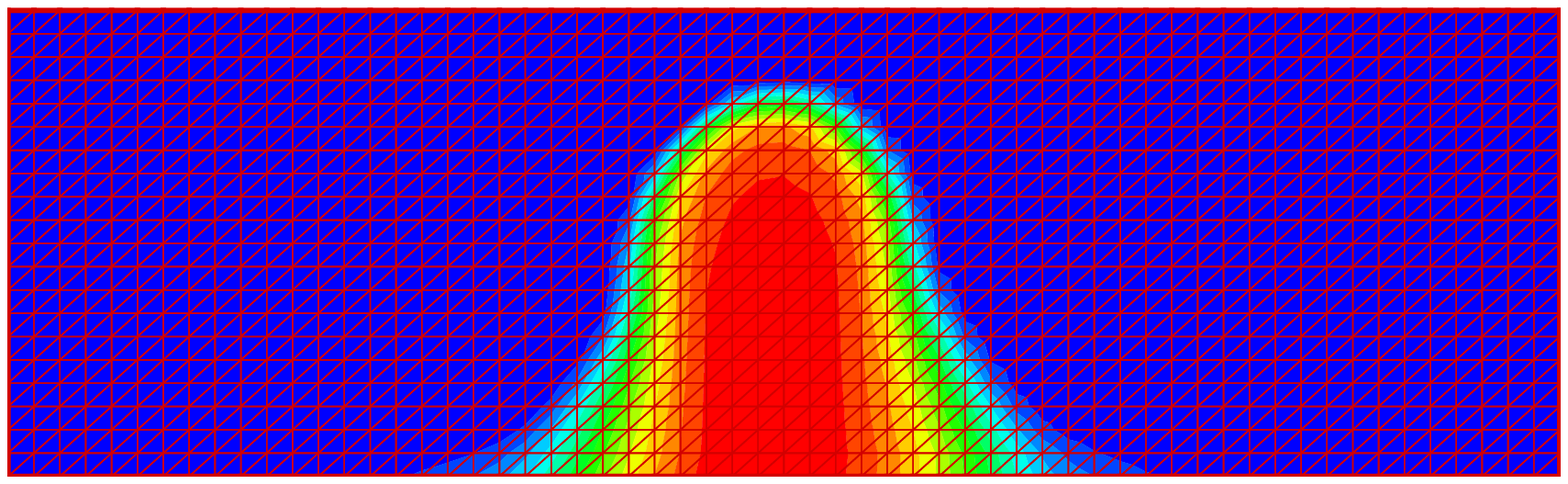} &
            \includegraphics[width=0.47\textwidth]{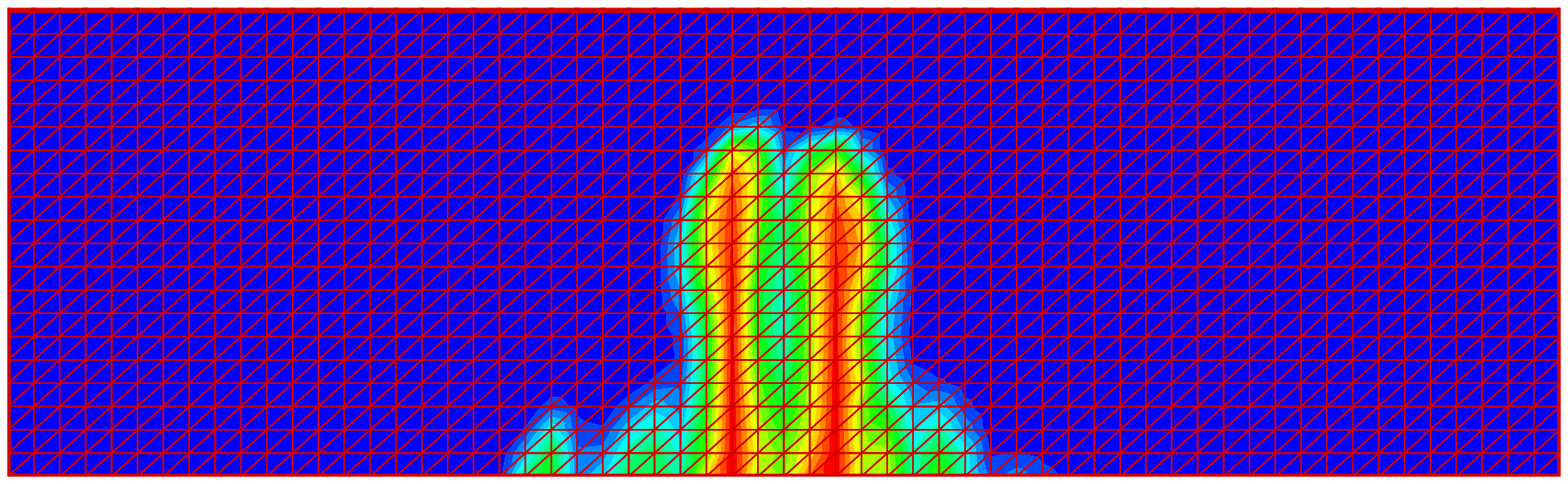} \\
            (a) & (b)
        \end{tabular}
        \end{center}
    \caption{Damage contours for two meshes (a) with gradient effects ($c\ne0$) and
            (b) without gradient effects ($c=0$).}
    \label{fig:meshes}
    \end{figure}

\begin{figure}
        \psfrag{x}{$u$}
        \psfrag{y}{$P$}
        \psfrag{coarseaaaaaa}{\tiny coarse $c=0$}
        \psfrag{caorsecaaaaa}{\tiny coarse $c \ne 0$}
        \psfrag{fineaaaaaaaa}{\tiny fine $c=0$}
        \psfrag{finecaaaaaaa}{\tiny fine $c \ne 0$}
        \center \includegraphics{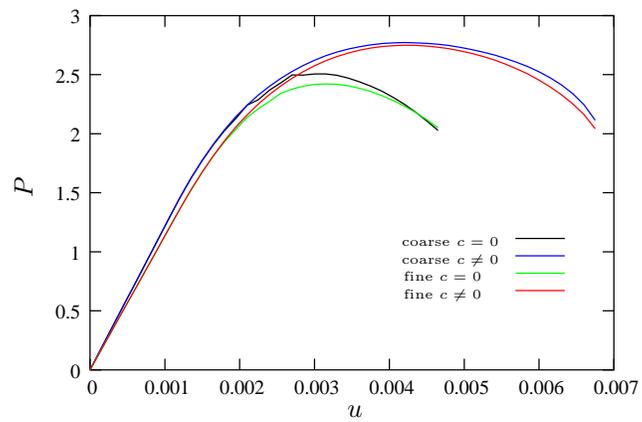}
    \caption{Load--displacement responses for two different meshes.}
    \label{fig:pd}
    \end{figure}

    The three-point bending test is performed for two different meshes with
$c \ne 0$ and $c=0$. The adopted geometry for the three-point bending test is
shown in Figure~\ref{fig:3bend}, and the adopted material parameters are:
Young's modulus $E=20 \times 10^{4}$~MPa, Poisson's ratio $\nu=0$, $\kappa_{0} =
1 \times 10^{-4}$, and  $\kappa_{c} = 1.25 \times 10^{-2}$. For
gradient-dependent simulations, $c=8 \times 10^{-2}$~mm.
Computations are stopped when damage reaches unity at any point in the mesh,
and damage development at the supports is prevented. The computed damage
contours for the four cases are shown in Figure~\ref{fig:meshes}.
From the damage contours, it is clear that the computed results are similar for
the two meshes with $c \ne 0$. In the absence of regularising effects ($c=0$),
the result is clearly affected by the discretisation.
    The load--displacement responses for the various cases are shown
in Figure~\ref{fig:pd}. Recall that a computation is halted when damage reaches
unity at a material point.
For the gradient-dependent case, the responses for the
two meshes are similar. For the case $c=0$, the responses are also similar,
which is somewhat in contrast to what is normally expected for a strain
softening problem. The responses are similar in this case due to the spurious
development of two cracks for the finer mesh (in contrast to the single main
crack for the coarse mesh). This is evident from the damage contours in
Figure~\ref{fig:meshes}.

\section{Conclusions}
\label{sect5}
    A discontinuous Galerkin formulation for a strain gradient-dependent damage
model has been investigated for a range of different finite elements. The model
allows the numerical solution of a continuum problem which would classically
require $C^{1}$ interpolations
with a simple $C^{0}$ or even discontinuous basis. Examples demonstrate robust
performance for a range of polynomial orders and degrees of continuity of the
interpolation functions, and are supported by rigorous error analysis.
Specifically, lower-order interpolations perform well and are relatively simple
to construct. The convergence properties of the proposed method have been
examined for the elastic case, for which the observed rates are consistent with
the theoretically predicted rates. The formulation has been observed
numerically to converge also
for damage problems. Finally, the formulation was applied successfully to a
two-dimensional problem. While the approach is promising, several issues
remain. Difficulties which must be resolved for other gradient models
include the effective imposition of boundary conditions on the fixed
boundary of a body, and at moving boundaries internal to a body. The development
of
thermodynamically consistent models would assist in this sense, as the
higher-order kinematic gradients have a natural partner in the energetic sense.

\section*{Acknowledgements}
    LM and FU acknowledge the support of University of Bologna, GNW acknowledges
the support of the Netherlands Technology Foundation (STW), and KG acknowledges
support from the US National Science Foundation by way of grant no. CMS0087019,
and from Sandia National Laboratory. The support from Sandia National Laboratory
includes a Presidential Early Career Award.

\bibliography{mybib}
\bibliographystyle{elsart-harv}

\end{document}